\documentclass{evn2004}
\usepackage{txfonts}
\begin{document}
\title{Spitzer 24 $\mu$m imaging of Faint Radio Sources in the FLSv:
 a new radio-loud, Mid-IR/optically obscured population?}

   \author{M. Orienti\inst{1,2}
          \and
          M.A. Garrett\inst{3}
     \and
     C. Reynolds\inst{3}
     \and
     R. Morganti\inst{4}
          }

   \institute{Dipartimento di Astronomia, Universit\`{a} di Bologna, Via Ranzani 1, I-40127 Bologna, Italy
         \and
             Istituto di Radioastronomia - CNR, via Gobetti 101, I-40129 Bologna, Italy
          \and
     Joint Institute for VLBI in Europe, Postbus 2, 7990 AA, Dwingeloo, The Netherlands
     \and
      Netherlands Foundation for Research in Astronomy, Postbus 2, 7990 AA, Dwingeloo, The Netherlands
             }
           
           \abstract{ Data from the Spitzer Space Telescope (the First
             Look Survey - FLS) have recently been made public. We have
             compared the 24 $\mu$m images with very deep WSRT 1.4 GHz
             observations (Morganti et al. 2004), centred on the FLS
             verification strip (FLSv).  Approximately 75\% of the radio
             sources have corresponding 24 $\mu$m identifications. Such
             a close correspondence is expected, especially at the
             fainter radio flux density levels, where star forming
             galaxies are thought to dominate both the radio and mid-IR
             source counts. Spitzer detects many sources that have no
             counter-part in the radio. However, a significant fraction
             of radio sources detected by the WSRT ($\sim 25$\%) have no mid-IR
             identification in the FLSv (implying a 24 $\mu$m flux
             density $\leq$ 100 $\mu$Jy). The fraction of radio sources
             without a counterpart in the mid-IR appears to increase
             with increasing radio flux density, perhaps indicating
             that some fraction of the AGN population may be detected
             more readily at radio than Mid-IR wavelenghts.
             We present
             initial results on the nature of the radio sources without
             Spitzer identification, using data from various
             multi-waveband instruments, including the publicly
             available R-band data from the Kitt Peak 4-m telescope.  }

\titlerunning{Spitzer 24 $\mu$m imaging of Faint Radio Sources in the FLSv}
\maketitle
%

\section{Introduction}
Deep radio surveys (S $\leq$ 1 mJy) have clearly indicated the
emergence of a new population of radio sources at mJy and sub-mJy
levels. At flux densities in excess of 1 mJy, the source counts are
dominated by AGN, in which the energy mechanism is believed to be
accretion of matter onto a supermassive black hole. Several class of
object have been invoked to explain the steep rise in the integral
radio source counts at faint sub-mJy levels: star forming galaxies,
similar to M~82 and Arp~220 (Rowan-Robinson et al. 1993); low-luminosity AGN
like M~84, and strongly evolving spirals (Condon 1989). \\
The fact that
the locally derived far-IR/radio correlation (e.g. Helou \& Bicay 1993)
also applies to the vast majority of the faint (and cosmologically
distant) radio source population (Garrett 2002), strongly supports the
idea that star forming galaxies begin to dominate the microJy radio source
population. \\
The recently launched {\it Spitzer Space Telescope}, is an order of
magnitude more sensitive than previous infrared-telescopes, providing
an important opportunity to constrain the nature of the sub-mJy radio
source population.
The First Look Survey (FLS) was the first survey
undertaken by Spitzer. In particular, a small but deeper subset of the
survey is focused on an area of 0.26 square degrees (the verification
strip or FLSv), reaching completely unexplored ($3\sigma$) sensitivity
level of $\sim$ 0.08 $\mu$Jy at 24 $\mu$m (Marleau et al. 2004). \\
In this paper we compare the deep (1$\sigma$ rms noise-level $\sim$
8.5 $\mu$Jy) WSRT radio image of the FLSv field
(Morganti et al. 2004) with the recent Spitzer 24 $\mu$m public images
of the same field (Fig. 1).\\



\section{The samples}

   \begin{figure}
   \centering
   \vspace{7.0cm}
   \includegraphics{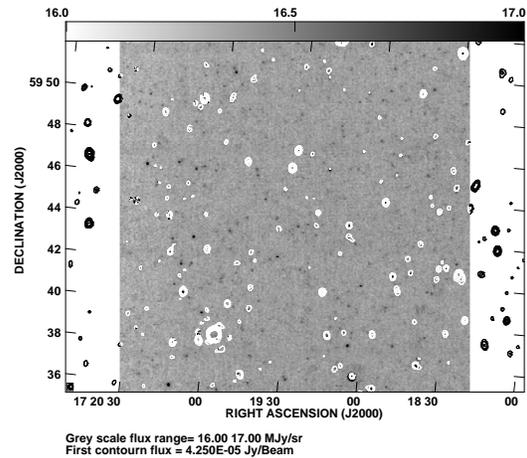}
   \caption{An example of Spitzer MIPS-24 image with the WSRT radio
   contours superimposed. The first contour is 5$\sigma$ = 42.5
   $\mu$Jy/beam, contour levels increase by a factor 2.
            \label{fig:thomas}
           }
    \end{figure}
    We have extracted a catalogue of sources observed by Spitzer's Multiband
    Imaging Photometer at 24 $\mu$m (MIPS-24), from the Post-Basic
    Calibrated Data (PBCD), using the Starfinder code (Diolaiti et al.
    2000). It should be noted that since the WSRT observations cover a
    bigger area than the FLSv field, only 389 sources of the 1048
    sources detected in
    the WSRT catalogue are located within the FLSv region. We
    identify two distinct samples from the FLSv radio catalogue: 
\begin{itemize}
\item Sample I: 292 radio sources with clear MIPS-24 identifications,
  comprising $\sim 75$\% of the complete FLSv radio sample; 
\item Sample II: 97 radio sources {\it without} MIPS-24
  identifications, comprising $\sim 25$\% of the complete FLSv radio sample.
\end{itemize}

\noindent
Both samples were cross-correlated with the optical R-band
FLS catalogue from the Kitt Peak 4-m telescope (Fadda et al. 2004). Although
the optical catalogue is estimated to be 50\% complete at R=24.5
(Vega), in both samples we find $\sim$ 20\% of radio sources without
optical identification.  

\section{Results}

Although Spitzer is able to pinpoint the sub-mm SCUBA population
(Frayer et al. 2004), there is a significant fraction of radio sources
($\sim$ 25\%) which have no MIPS-24 counterparts (Sample II). 
The two radio
samples have a different radio flux density distribution: Sample I is
dominated by the faintest radio sources, with flux densities typically
$\leq$ 300 $\mu$Jy.  Sample II appears to comprise the brighter radio
sources typically $\geq$ 1~mJy.

   \begin{figure}
   \centering
   \vspace{6.5cm}
   \includegraphics{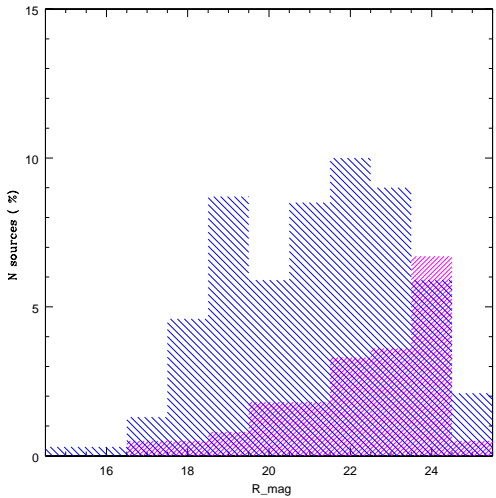}
   \caption{The R-band magnitude distribution of the radio sources in the
   FLSv field. The radio sources with MIPS-24 counterparts are indicated in
   blue, while the radio sources without MIPS-24 identification are in
   magenta.
            \label{fig:thomas}
           }
    \end{figure}
    Figure 2 shows that the R-band magnitude distributions of the two radio
    source samples are also quite different.  In particular, while 53\%
    of the radio sources with MIPS-24 identification (Sample I) have
    optical counterparts brighter than R=22.5, this figure is only 35\%
    for Sample II. These results suggest that the two samples are
    dominated by two different source populations. This is in
    agreement with the hypothesis that the mJy population is dominated
    by the faint tail of the AGN population, while star forming
    galaxies dominate at sub-mJy and microJy radio flux density levels
    (Prandoni et al. 2001; Richard 2000). Our results suggest that the
    radio sources without MIPS-24 identification (Sample II) are likely
    to be dominated by distant low-luminosity AGN. A study of the SED of
    various class of objects projected to various redshifts also
    supports this hypothesis. For example, an Ultra-luminous IR Galaxy
    like Arp 220 is detectable to z $\sim$ 0.7 with both the WSRT and
    Spitzer (see Fig.~3). However, a low-luminosity AGN can be detected
    up to z $\sim$ 0.7 by WSRT but only to z $\sim$ 0.15 by Spitzer.
    Another possible explanation is related to the mass and temperature
    of the dust in the host galaxy. A star forming galaxy with the same
    dust mass of Arp 220, but with a lower temperature (e.g. $\leq$ 30
    K), is only detectable to z $\sim$ 0.1 by MIPS-24. 
   \begin{figure}
   \centering
   \vspace{6.5cm}
   \includegraphics{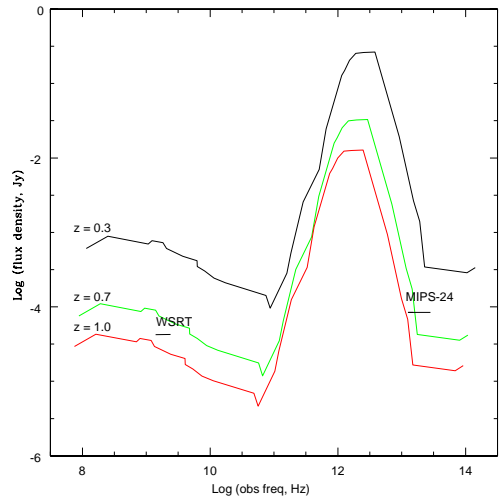}
   \caption{The SED of Arp 220 (at radio, sub-mm and FIR frequencies)
   projected to various redshifts (z=0.3, 0.7, 1.0). The 5$\sigma$ and
   3$\sigma$ detection threshold for both WSRT at 1.4 GHz and Spitzer
   at 24 $\mu$m respectively (solid lines), are presented.
            \label{fig:thomas}
           }
    \end{figure}
VLBI, sub-mm and X-ray observations will be crucial in order to
further constrain the nature of this class of radio source not detected by
Spitzer at 24 micron.  
\begin{acknowledgements}
This work is based in part on observations made with the {\it Spitzer
  Space Telescope}, which is operated by the JPL, California Institute
  of Technology, under NASA contract 1407. The National Optical
  Astronomy Observatory (NOAO) is operated by the Association of
  Universities for Research in Astronomy (AURA), Inc. under
  cooperative agreement with the National Science Fundation.
\end{acknowledgements}

\end{document}